\documentclass[journal]{IEEEtran}
\usepackage[cmex10]{amsmath}
\usepackage{amssymb}
\usepackage{amscd}
\newtheorem{theorem}{Theorem}
\newtheorem{lemma}[theorem]{Lemma}

\newtheorem{proposition}[theorem]{Proposition}

\newtheorem{example}[theorem]{Example}
\newtheorem{remark}[theorem]{Remark}

\ifCLASSINFOpdf
\else
\fi
\begin{document}
%
\title{Improved Linear Programming Bounds on Sizes of Constant-Weight Codes}
%
%
%

\author{Byung~Gyun~Kang, 
        Hyun~Kwang~Kim, 
        and~Phan~Thanh~Toan 
\thanks{B.G.~Kang, H.K.~Kim, and P.T.~Toan are with the Department of Mathematics, Pohang University of Science and Technology, Pohang 790-784, Republic of Korea.}
}

\maketitle

\begin{abstract}
Let $A(n,d,w)$ be the largest possible size of an $(n,d,w)$
constant-weight binary code. By adding new constraints to Delsarte
linear programming, we obtain twenty three new upper bounds on
$A(n,d,w)$ for $n \leq 28$. The used techniques allow us to give a
simple proof of an important theorem of Delsarte which makes
linear programming possible for binary codes.
\end{abstract}


%
\IEEEpeerreviewmaketitle

\section{Introduction}
%
%
%
%

\IEEEPARstart{L}{et} $\mathcal{F} = \{0,1\}$ and let $n,d,w$ be positive integers.
The {\em (Hamming) distance} between two vectors in $\mathcal{F}^n$
is the number of positions where they differ. An $(n,d,w)$
{\em constant-weight code} is a subset $\mathcal C$ of
$\mathcal{F}^n$ such that every vector of $\mathcal C$ has exactly
$w$ ones and such that distance between any two vectors in
$\mathcal C$ is at least $d$. Given $n,d,w$, denote $A(n,d,w)$
the largest possible size of an $(n,d,w)$ constant-weight code
$\mathcal C$. In general, it is difficult to find the exact values
of $A(n,d,w)$. However, many methods have been developed to find
lower bounds and upper bounds for $A(n,d,w)$. In this paper, we
only deal with the problem improving upper bounds on $A(n,d,w)$.
For lower bounds on $A(n,d,w)$, the readers may refer to
\cite{avz2, bbm, b, bsss, gs}.

We give a brief history of improvements of upper bounds on
$A(n,d,w)$. In 1977, first tables of upper bounds on $A(n,d,w)$
appeared in \cite{ms} for $n \leq 24$. Later, in 1978, they are
updated in \cite{bbm}. More updates appeared in 1987 \cite{h}. In
2000, Agrell, Vardy, and Zeger made very nice improvements on
upper bounds on $A(n,d,w)$ for $n \leq 28$ \cite{avz}. Many of
these upper bounds were obtained by adding new constraints to
Delsarte linear programming. Five years later, in 2005, Schrijver
also obtained great improvements on upper bounds on $A(n,d,w)$ by
using Terwilliger algebra and semidefinite programming \cite{s}.
And by computer-aided approach, in 2010, \"{O}sterg\aa rd
classified up to equivalence optimal constant-weight codes for
small $n$. Several upper bounds on $A(n,d,w)$ were also obtained
by this approach \cite{o}.

In this paper, we show that the distance distribution of a
constant-weight code satisfies certain linear inequalities. And by
adding these new constraints to linear programming, we obtain
twenty three new upper bounds on $A(n,d,w)$ for $n \leq 28$. The
remaining of the paper is organized as follows. In section \ref{ss2}, we
recall the Delsarte linear programming for constant-weight codes.
Next, in section \ref{ss3} and \ref{ss4}, we show two types of constraints that
are added to improved upper bounds on $A(n,d,w)$. The second type
of constraints, in many cases, helps reduce known upper bounds
on $A(n,d,w)$ by $1$. The techniques which are used to obtain these constraints allow us in section \ref{ss5}
to give a simple proof of well known Delsarte inequalities which
make linear programming possible for binary codes.



\section{Delsarte linear programming bounds on sizes of constant-weight codes}\label{ss2}
\subsection{Upper bounds on $A(n,d,w)$}

The following theorem shows easy properties of $A(n,d,w)$ which
can be found in \cite{ms}.
\begin{theorem}
\begin{equation}\label{mm}
A(n,d,w) = A(n,d+1,w), \quad \mbox{ if } d \mbox{ is odd,}
\end{equation}
\begin{equation}\label{mh}
A(n,d,w) = A(n,d,n-w),
\end{equation}
\begin{equation}\label{mb}
A(n,2,w) = \left(n \atop w \right),
\end{equation}
\begin{equation}\label{mbo}
A(n,2w,w) = \left\lfloor \frac{n}{w} \right\rfloor,
\end{equation}
\begin{equation}\label{mn}
A(n,d,w) = 1, \quad \mbox{ if } 2w<d.
\end{equation}
\end{theorem}

By (\ref{mm}) and (\ref{mb}), we can always assume that $d$ is even and $d \geq 4$. Also, by (\ref{mh}), (\ref{mbo}), and (\ref{mn}), we can assume that $d < 2w \leq n$. From now on, $n, d,$ and $w$ are assumed to satisfy these conditions. And a constant-weight code means an $(n,d,w)$ constant-weight code.

The following theorem of Johnson in some cases still gives best
known upper bounds on sizes of constant-weight codes.

\begin{theorem} \label{Johns} (Johnson).
\begin{eqnarray}
A(n,d,w) \leq \left \lfloor \frac{n}{w} A(n-1, d, w-1) \right \rfloor,
\end{eqnarray}
\begin{eqnarray}
A(n,d,w) \leq \left \lfloor \frac{n}{n-w} A(n-1, d, w) \right
\rfloor.
\end{eqnarray}
\end{theorem}

Let $u, v$ be two vectors in $\mathcal{F}^n$. If the distance
between $u$ and $v$ is $i$, then we write $d(u,v) = i$. Let
$\mathcal C$ be a constant-weight code. The {\em distance
distribution} $\{A_i\}_{i=0}^n$ of $\mathcal C$ is define by
\begin{equation}
A_{i} = \frac{1}{|\mathcal C|} \sum_{c \in \mathcal C} |S_{i}(c)|
\end{equation}
for $i = 0,1, \ldots, n$, where
\begin{eqnarray}
S_{i}(c) = \{u \in \mathcal C \mid d(u,c) = i\},
\end{eqnarray}
the set of all codewords $u$ in $\mathcal C$ at distance $i$ from $c$.
\begin{remark}{\rm
By definition, $A_0 = 1$ and $A_{j} = 0$ whenever $0<j<d$ or $2w <
j$ or when $j$ is odd. Hence, the possible nonzero $A_j$ are $A_0,
A_{d}, A_{d+2}, \ldots, A_{2w}$, which are
\begin{eqnarray}
A_0 \mbox{ and } A_{2i},  \quad i = d/2, \ldots, w.
\end{eqnarray}
Since $A_0 = 1$, we can consider $\{A_{2i}\}_{i = d/2}^w$ (sometimes we just write $\{A_{2i}\}$ for
short) as the distance distribution of $\mathcal C$. Note that if
$\mathcal C$ is a constant-weight code with distance distribution
$\{A_{2i}\}$, then
\begin{eqnarray}
|\mathcal C| = \left(\sum_{i=d/2}^w A_{2i}\right) + A_0 = \left(\sum_{i=d/2}^w A_{2i}\right) + 1.
\end{eqnarray}
}
\end{remark}

\begin{theorem}\label{dtr} (Delsarte).
If $\{A_{2i}\}$ is the distance distribution of an $(n,d,w)$ constant-weight code, then for $k = 1, 2, \ldots, w$,
\begin{eqnarray}
\sum_{i=d/2}^w q(k,i,n,w)A_{2i} \geq -1,
\end{eqnarray}
where
\begin{eqnarray}
q(k,i,n,w) = \frac{\sum_{j=0}^i (-1)^j \left( k \atop j \right) \left(w-k \atop i-j \right) \left( n-w-k \atop i - j \right)}{\left(w \atop i \right) \left(n-w \atop i \right)}.
\end{eqnarray}
\end{theorem}

The original version of linear programming bound is stated as follows.

\begin{theorem}
\begin{eqnarray}
A(n,d,w) \leq \left \lfloor \max \sum_{i=d/2}^w A_{2i} \right \rfloor + 1,
\end{eqnarray}
where the maximum is taken over all $(A_d, A_{d+2}, \ldots, A_{2w})$ satisfying $A_{2i} \geq 0$ for $i = d/2, \ldots, w$ and satisfying the constraints in Theorem \ref{dtr}.
\end{theorem}

\subsection{Some improvements}
If more constraints are added to this linear programming, then
better upper bounds on $A(n,d,w)$ may be obtained. One way to do
this is using upper bounds on sizes of doubly-constant-weight
codes. Let $w_1, n_1, w_2, n_2$ be nonnegative integers. A $(w_1,
n_1, w_2, n_2,d)$ {\em doubly-constant-weight code} is a subset
$\mathcal C$ of $\mathcal{F}^{n_1+n_2}$ such that each vector in
$\mathcal C$ has exactly $w_1$ ones on the first $n_1$ coordinates
and exactly $w_2$ ones on the last $n_2$ coordinates and such that
distance between any two vectors in $\mathcal C$ is at least $d$.
Denote $T(w_1,n_1,w_2,n_2,d)$ the largest possible size of a
$(w_1,n_1,w_2,n_2,d)$ doubly-constant-weight code. Some elementary
facts on $T(w_1,n_1,w_2,n_2,d)$ are as follows.

\begin{theorem}
\begin{eqnarray}
T(w_1,n_1,w_2,n_2,d) = T(w_2,n_2,w_1,n_1,d),
\end{eqnarray}
\begin{eqnarray}\label{tru}
T(w_1,n_1,w_2,n_2,d) = T(n_1-w_1,n_1,w_2,n_2,d),
\end{eqnarray}
\begin{eqnarray}
T(0,n_1,w_2,n_2,d) = A(n_2,d,w_2),
\end{eqnarray}
\begin{eqnarray}
T(w_1,n_1,w_2,n_2,2) = \left(n_1 \atop w_1 \right) \left(n_2 \atop w_2  \right),
\end{eqnarray}
\begin{eqnarray}
T(w_1,n_1,w_2,n_2,2w_1+2w_2) = \min_{i=1,2} \left\{\left \lfloor \frac{n_i}{w_i} \right \rfloor \right\},
\end{eqnarray}
\begin{eqnarray}
\nonumber T(w_1,n_1,w_2,n_2,d) = T(w_1,n_1,w_2,n_2,d+1), \\
\mbox{ if } d
\mbox{ is odd},
\end{eqnarray}
\begin{eqnarray}
T(w_1,n_1,w_2,n_2,d) = 1, \mbox{ if } 2w_1 + 2w_2 < d.
\end{eqnarray}
\end{theorem}
Best known upper bounds on $T(w_1,n_1,w_2,n_2,d)$ can be found
at \cite{avz2}. The following inequalities are well known. We
gives the proof here for the completeness.

\begin{lemma}\label{lemmaa1}
Let $\mathcal C$ be  a constant-weight code with distance distribution $\{A_{2i}\}.$ Then for each $c \in \mathcal C$ and each $i$,
\begin{eqnarray} \label{s2i2j}
|S_{2i}(c)| \leq T(i,w, i, n-w, d).
\end{eqnarray}
\end{lemma}
\begin{proof}
Let $u \in S_{2i}(c)$. By reordering the coordinates, we may assume that
\begin{eqnarray}
c = & \overbrace{1 \ldots 1 1 \ldots 1}^{w} & \overbrace{0 \ldots 0 0 \ldots 0}^{n - w} \\
u = & {1 \ldots 1} \underbrace{0 \ldots 0}_{i} & \underbrace{1 \ldots 1}_i  {0 \ldots 0}
\end{eqnarray}
Since $d(u,c)=2i$, $u$ must have exactly $i$ zeros on the first $w$ coordinates and exactly $i$ ones on the last $n -w$ coordinates. It follows, by (\ref{tru}), that
\begin{eqnarray}
\nonumber |S_{2i}(c)| & \leq &  T(w-i,w, i, n-w, d)\\
 & = & T(i,w, i, n-w, d).
\end{eqnarray}
\end{proof}

Lemma \ref{lemmaa1} leads to the following well known constraints which can be added to the linear programming.
\begin{proposition} \label{t1}
Let $\mathcal C$ be  a constant-weight code with distance distribution $\{A_{2i}\}.$ Then for each $i$,
\begin{eqnarray}
A_{2i} \leq T(i,w, i, n-w, d).
\end{eqnarray}
\end{proposition}

\begin{proof}
By Lemma \ref{lemmaa1}, for each $c \in \mathcal C$,
\begin{eqnarray}
|S_{2i}(c)| \leq T(i,w, i, n-w, d).
\end{eqnarray}
Taking sum over all $c \in \mathcal C$, we get
\begin{eqnarray}
\sum_{c \in \mathcal C}|S_{2i}(c)| \leq  |\mathcal C| T(i,w, i, n-w, d),
\end{eqnarray}
which means
\begin{eqnarray}
A_{2i} \leq T(i,w, i, n-w, d).
\end{eqnarray}
\end{proof}

\begin{remark}{\rm
In Proposition \ref{t1}, the exact values of $T(i,w,i,n-w,d)$ may not be known. However, we can replace them by upper bounds of $T(i,w,i,n-w,d)$ taken from the tables at \cite{avz2}.}
\end{remark}

More improvements on linear programming bounds on $A(n,d,w)$ can be found at \cite{avz}.

\section{Improved linear programming bounds on $A(n,d,w)$} \label{ss3}

We now construct the first type of constraints on $\{A_{2i}\}_{i=d/2}^w$ that
will be added to the linear programming to improve upper bounds on
$A(n,d,w)$. This type of constraints is similar to one in \cite{avz}.
Let $\mathcal C$ be a constant-weight code with
distance distribution $\{A_{2i}\}_{i=d/2}^w$. For convenience, we
denote $H = \{d/2, d/2+1, \ldots, w\}$. For each $i \in H$, we let
$V_{i}$ be the set of all vectors $u$ of $\mathcal{F}^n$ such that
$u$ has exactly $i$ ones on the first $w$ coordinates and exactly
$i$ ones on the last $n-w$ coordinates. And for $i \not = j$ in
$H$, we define
\begin{eqnarray}
m_{i,j}= \max\{d(u,v) \mid u \in V_{i}, v \in V_{j}\}.
\end{eqnarray}
This $m_{i,j}$ can be calculated easily.

\begin{proposition}
For $i$ and $j$ in $H$,
\begin{eqnarray}
m_{i,j} = a + b,
\end{eqnarray}
where
\begin{eqnarray}
a = \left\{ \begin{array}{ll}
i+j & \mbox{if } i+j \leq w \\
i+j - 2(i+j-w) & \mbox{if } i+j>w
\end{array} \right.,
\end{eqnarray}
and
\begin{eqnarray}
b = \left\{ \begin{array}{ll}
i+j & \mbox{if } i+j \leq n-w \\
i+j - 2[i+j-(n-w)] & \mbox{if } i+j>n-w
\end{array} \right..
\end{eqnarray}

\end{proposition}

\begin{proof}
Straightforward.
\end{proof}

From now on, for $i \in H$, $P_{i}$ always denote an integer such that
\begin{eqnarray}\label{s2i2jt}
P_{i} \geq T(i,w,i,n-w,d).
\end{eqnarray}
Hence by (\ref{s2i2j}), for every $c \in \mathcal C$, we always have
\begin{eqnarray}\label{s2i2jp}
|S_{2i}(c)| \leq P_{i}.
\end{eqnarray}

\begin{lemma} \label{ll2}
Let $\mathcal C$ be a constant-weight code with distance distribution $\{A_{2i}\}$. If $i \not = j$ are in $H$ such that $m_{i,j}<d$, then
\begin{eqnarray}
\frac{|S_{2i}(c)|}{P_i} + \frac{|S_{2j}(c)|}{P_j} \leq 1
\end{eqnarray}
for each $c \in \mathcal C$.
\end{lemma}

\begin{proof}
The proof follows from (\ref{s2i2jp}) and the following claim.

{\it Claim. Either $|S_{2i}(c)| =0$ or $|S_{2j}(c)| = 0$.} Suppose on the contrary that $|S_{2i}(c)| \geq 1$ and $|S_{2j}(c)| \geq 1$. Then choose any $u \in S_{2i}(c)$ and $v \in S_{2j}(c)$. Then $u+c$ belongs to $V_{i}$ and $v+c$ belongs to $V_{j}$. By definition of $m_{i,j}$, $d(u+c,v+c) \leq m_{i,j}$. Thus, $d(u,v) = d(u+c,v+c) \leq m_{i,j} < d$ and hence $u = v$ which is a contradiction since $i \not = j$.
\end{proof}

\begin{proposition} \label{l2}
Let $\mathcal C$ be a constant-weight code with distance distribution $\{A_{2i}\}$. If $i \not = j$ are in $H$ such that $m_{i,j}<d$, then
\begin{eqnarray}
\frac{A_{2i}}{P_i} + \frac{A_{2j}}{P_j} \leq 1.
\end{eqnarray}
\end{proposition}

\begin{proof}
For each $c \in \mathcal C$, by Lemma \ref{ll2},
\begin{eqnarray}
\frac{|S_{2i}(c)|}{P_i} + \frac{|S_{2j}(c)|}{P_j} \leq 1.
\end{eqnarray}
Taking sum over all $c \in \mathcal C$, we get
\begin{eqnarray}
|\mathcal C| \frac{A_{2i}}{P_i} + |\mathcal C| \frac{ A_{2j}}{P_j} \leq |\mathcal C|.
\end{eqnarray}
Hence,
\begin{eqnarray}
\frac{A_{2i}}{P_i} + \frac{A_{2j}}{P_j} \leq 1.
\end{eqnarray}
\end{proof}



Proposition \ref{l2} can be generalized as follows.
\begin{proposition} \label{dl4}
Let $\mathcal C$ be a constant-weight code with distance distribution $\{A_{2i}\}$. If $H_1$ is a subset of $H$ such that $|H_1| \geq 2$ and such that $m_{i,j}<d$ for any $i \not = j$ in $H_1$, then
\begin{eqnarray}
\sum_{i \in H_1} \frac{A_{2i}}{P_i} \leq 1.
\end{eqnarray}
\end{proposition}

\begin{proof}
As in the proof of Lemma \ref{ll2}, we have, for each $c \in \mathcal C$,
\begin{eqnarray}
\sum_{i \in H_1} \frac{|S_{2i}(c)|}{P_i} \leq 1.
\end{eqnarray}
This is true because if $|S_{2i}(c)| \geq 1$, then $|S_{2j}(c)| = 0$, for all $j \in H_1$ different from $i$. As in the proof of Proposition \ref{l2}, we take sum over all $c \in \mathcal C$ and get the desired result.
\end{proof}


We now consider the case $m_{i,j} = d$ for some $i \not = j \in
H$. We first define two number $w(i,j,t)$ and $n(i,j,t)$ when $i,
j, t$ are given. For integers $i, j, t$, we let
\begin{eqnarray}
w(i,j,t) = |i+j-t|,
\end{eqnarray}
the absolute value of $i+j-t$, and let
\begin{eqnarray}
n(i,j,t) = \left\{
\begin{array}{ll}
i & \mbox{if } t < i+j\\
t-i & \mbox{if } t \geq i+j
\end{array}
\right..
\end{eqnarray}

\begin{lemma} \label{lm7}
Suppose $i \not = j$ are in $H$ such that $m_{i,j} = d$. Let $c \in C$. If $|S_{2i}(c)| \geq 1$, then
\begin{eqnarray}
|S_{2j}(c)| \leq T(w_1, n_1, w_2, n_2,d),
\end{eqnarray}
where
\begin{eqnarray}
w_1 & = & w(i,j,w), \\ n_1 & = & n(i,j,w),\\
w_2 & = & w(i,j,n-w),  \\n_2  & = & n(i,j,n-w).
\end{eqnarray}
\end{lemma}

\begin{proof}
Let $u \in  S_{2i}(c)$. If $S_{2j}(c)$ is empty then there is nothing to prove. Hence, we assume that $|S_{2j}(c)| \geq 1$. Let $v \in S_{2j}(c)$. Since $u+c \in V_i$ and $v+c \in V_j$, we have $d  \leq  d(u,v) = d(u+c,v+c)  \leq m_{i,j} = d$. Thus, $d(u+c, v+c) = m_{i,j}$. We may write
\begin{eqnarray}
u + c & = &\underbrace{\overbrace{1 \cdots 1}^i 0 \cdots 0}_{w} \underbrace{\overbrace{1 \cdots 1}^i 0 \cdots 0}_{n-w}.
\end{eqnarray}

{\em Case $1$.} $w \geq i + j$.
\begin{eqnarray}
\begin{array}{lcccc}
u+c & = & \overbrace{1 \cdots 1}^i & 0 \cdots 0 & 0 \cdots 0 | \cdots \\
v+c & = & 0 \cdots 0 & \underbrace{0 \cdots 0}_{w-i-j} & \underbrace{1 \cdots 1}_j | \cdots\\
\end{array}.
\end{eqnarray}
On the first $w$ coordinates, the $j$ ones of $v+c$ are free to run over $w - i$ coordinates, i.e., no ones of $v+c$ are allowed to be on the first $i$ coordinates, since $d(u+c,v+c) = m_{i,j}$. Equivalently, $w - i - j$ zeros of $v + c$ on the first $w$ coordinates are free to run over $w - i$ coordinates and the other zeros of $v+c$ are fixed. This means we are allowed to choose $w - i - j = w(i,j,w) = w_1$ coordinates from $w - i = n(i,j,w) = n_1$ coordinates.

{\em Case $2$.} $w < i+j$.
\begin{eqnarray}
\begin{array}{lcccc}
u+c & = & {1 \cdots 1} & 1 \cdots 1 & \overbrace{0 \cdots 0}^{w-i} | \cdots \\
v+c & = & \underbrace{0 \cdots 0}_{w-j} & \underbrace{1 \cdots 1}_{i+j-w} & {1 \cdots 1} | \cdots \\
\end{array}.
\end{eqnarray}
In this case, only $i+j -w$ ones of $v+c$ on the first $w$ coordinates are free to run over the first $i$ coordinates and the other $w-i$ ones must be fixed since $d(u+c,v+c)=m_{i,j}$. This means we are allowed to choose $i+j-w = w(i,k,w)=w_1$ coordinates from $i = n(i,j,w)$ coordinates.

Therefore, in any cases, on the first $w$ coordinates of $v+c$, we are allowed to choose $w_1$ coordinates from $n_1$ coordinates.

Similarly, on the last $n-w$ coordinates, we are free to choose $w_2$ coordinates from $n_2$ coordinates.

The conclusion is that
\begin{eqnarray}
|S_{2j}(c)| \leq T(w_1, n_1, w_2, n_2,d).
\end{eqnarray}
\end{proof}

From now on, $P_{ji}$ always denote an integer such that
\begin{eqnarray} \label{muoi}
P_{ji} \geq T(w_1, n_1, w_2, n_2,d),
\end{eqnarray}
where $i, j, w_1, n_1, w_2, n_2$ are as in Lemma \ref{lm7}.

\begin{lemma} \label{l5'}
Let $\mathcal C$ be a constant-weight code with distance distribution $\{A_{2i}\}$. If $i \not = j$ are in $H$ such that $m_{i, j}=d$, then for each $c \in \mathcal C$,
\begin{eqnarray} \label{mot1}
\nonumber \frac{P_j - P_{ji}}{P_jP_{ij}} |S_{2i}(c)| + \frac{1}{P_j}|S_{2j}(c)| \leq 1, \\
\mbox{ if } \displaystyle \frac{P_{ij}}{P_{i}} + \frac{P_{ji}}{P_{j}} \geq 1,
\end{eqnarray}
\begin{eqnarray} \label{hai2}
\nonumber \frac{1}{P_i}|S_{2i}(c)| + \frac{P_{i} - P_{ij}}{P_{i}P_{ji}} |S_{2j}(c)| \leq 1, \\
\mbox{ if } \displaystyle \frac{P_{ij}}{P_{i}} + \frac{P_{ji}}{P_{j}} \geq 1,
\end{eqnarray}
\begin{eqnarray} \label{ba3}
\frac{1}{P_i}|S_{2i}(c)| + \frac{1}{P_j}|S_{2j}(c)| \leq 1, \quad \mbox{ if } \displaystyle \frac{P_{ij}}{P_{i}} + \frac{P_{ji}}{P_{j}} \leq 1.
\end{eqnarray}
\end{lemma}

\begin{proof}
Fix $c \in \mathcal C$. By (\ref{s2i2jp}),
\begin{eqnarray} \label{mbon}
|S_{2i}(c)| \leq P_{i} \mbox{ and } |S_{2j}(c)| \leq P_{j}.
\end{eqnarray}

By Lemma \ref{lm7} and (\ref{muoi}),
\begin{eqnarray} \label{mlam}
|S_{2i}(c)| \leq P_{ij}, \mbox{ if } |S_{2j}(c)| \geq 1, \\
|S_{2j}(c)| \leq P_{ji}, \mbox{ if } |S_{2i}(c)| \geq 1.
\end{eqnarray}

First, we prove (\ref{mot1}) by considering the following three cases.

{\em Case $1$.} $|S_{2i}(c)| = 0$. It is obvious by (\ref{mbon}).

{\em Case $2$.} $|S_{2i}(c)|  \geq 1$ and $|S_{2j}(c)|=0$. We need to show that
\begin{eqnarray}
(P_{j}-P_{ji}) |S_{2i}(c)| \leq P_{j}P_{ij}.
\end{eqnarray}
By hypothesis, $\displaystyle \frac{P_{ij}}{P_{i}} + \frac{P_{ji}}{P_{j}} \geq 1$. Thus, $ P_{i}P_{j} -P_{i}P_{ji} \leq P_{j}P_{ij}$ and hence
\begin{eqnarray}
(P_{j}-P_{ji}) |S_{2i}(c)| \leq (P_{j}-P_{ji}) P_{i} \leq P_{j}P_{ij}.
\end{eqnarray}

{\em Case $3$.} $|S_{2i}(c)|  \geq 1$ and $|S_{2j}(c)| \geq 1$.

\begin{eqnarray}
\nonumber \displaystyle \frac{P_j - P_{ji}}{P_jP_{ij}} |S_{2i}(c)| + \frac{1}{P_j}|S_{2j}(c)| & \leq & \displaystyle \frac{P_j - P_{ji}}{P_jP_{ij}} P_{ij} + \frac{1}{P_j}P_{ji} \\
\nonumber & = &\displaystyle 1- \frac{P_{ji}}{P_j} + \frac{P_{ji}}{P_j}\\
& = & 1.
\end{eqnarray}

By symmetry, (\ref{hai2}) follows.

Now, we prove (\ref{ba3}). By (\ref{mbon}), the proof is trivial if $|S_{2i}(c)|=0$ or $|S_{2j}(c)|=0$. Suppose $|S_{2i}(c)| \geq 1$ and $|S_{2j}(c)|\geq 1$. Then

\begin{eqnarray}
\frac{1}{P_i}|S_{2i}(c)| + \frac{1}{P_j}|S_{2j}(c)|  \leq  \displaystyle  \frac{1}{P_i} P_{ij} + \frac{1}{P_j} P_{ji} \leq  1.
\end{eqnarray}
\end{proof}

\begin{proposition} \label{dl6}
Let $\mathcal C$ be a constant-weight code with distance distribution $\{A_{2i}\}$. If $i \not = j$ are in $H$ such that $m_{i,j}=d$, then
\begin{eqnarray} \label{mot}
\frac{P_j - P_{ji}}{P_jP_{ij}} A_{2i} + \frac{1}{P_j}A_{2j} \leq 1, \quad \mbox{ if } \displaystyle \frac{P_{ij}}{P_{i}} + \frac{P_{ji}}{P_{j}} \geq 1,
\end{eqnarray}
\begin{eqnarray} \label{hai}
\frac{1}{P_i}A_{2i} + \frac{P_{i} - P_{ij}}{P_{i}P_{ji}} A_{2j} \leq 1, \quad \mbox{ if } \displaystyle \frac{P_{ij}}{P_{i}} + \frac{P_{ji}}{P_{j}} \geq 1,
\end{eqnarray}
\begin{eqnarray} \label{ba}
\frac{1}{P_i}A_{2i} + \frac{1}{P_j}A_{2j} \leq 1, \quad \mbox{ if } \displaystyle \frac{P_{ij}}{P_{i}} + \frac{P_{ji}}{P_{j}} \leq 1.
\end{eqnarray}
\end{proposition}

\begin{proof}
Follows from Lemma \ref{l5'}.
\end{proof}

Proposition \ref{dl6} can be improved as follows.

\begin{proposition}\label{dl7}
Let $\mathcal C$ be a constant-weight code with distance distribution $\{A_{2i}\}$. Suppose $H_1$ is a subset of $H$ satisfying the following properties.
\begin{itemize}
\item [(a)] $|H_1| \geq 2$.
\item [(b)] There exist $i \not = j$ in $H_1$ such that $m_{i,j} = d$.
\item [(c)] For any $k \not = l \in H_1$, if $k \not = i$ or $l \not = j$, then $m_{k,l}< d$.
\end{itemize}
Let $H_2 = H_1 \setminus \{i, j\}$. Then

\begin{eqnarray} \label{hmuoi}
\nonumber \frac{P_j - P_{ji}}{P_jP_{ij}} A_{2i} + \frac{1}{P_j}A_{2j} + \sum_{k \in H_2} \frac{A_{2k}}{P_k}\leq 1, \\
\mbox{ if } \displaystyle \frac{P_{ij}}{P_{i}} + \frac{P_{ji}}{P_{j}} \geq 1,
\end{eqnarray}

\begin{eqnarray} \label{hmot}
\nonumber \frac{1}{P_i}A_{2i} + \frac{P_{i} - P_{ij}}{P_{i}P_{ji}} A_{2j} + \sum_{k \in H_2} \frac{A_{2k}}{P_k} \leq 1, \\
 \mbox{ if } \displaystyle \frac{P_{ij}}{P_{i}} + \frac{P_{ji}}{P_{j}} \geq 1,
\end{eqnarray}

\begin{eqnarray}\label{hhai}
\sum_{k \in H_1} \frac{A_{2k}}{P_k} \leq 1,  \quad \mbox{ if }\displaystyle \frac{P_{ij}}{P_{i}} + \frac{P_{ji}}{P_j} \leq 1.
\end{eqnarray}
\end{proposition}

\begin{proof}
The proof follows from previous results and the following two facts.
\begin{itemize}
\item {\em Fact $1$:} If there exists $k \in H_2$ such that $|S_{2k}(c)|\geq 1$, then $|S_{2l}(c)|=0$ for all $l \in H_1$ different from $k$.
\item {\em Fact $2$:} If $|S_{2i}(c)| \geq 1$ or $|S_{2j}(c)| \geq 1$, then $|S_{2k}(c)|=0$ for all $k \in H_2$.
\end{itemize}
For example, to prove (\ref{hmuoi}), we fix $c \in \mathcal C$. Then, by Lemma \ref{l5'},
\begin{eqnarray}
\frac{P_j - P_{ji}}{P_jP_{ij}} |S_{2i}(c)| + \frac{1}{P_j}|S_{2j}(c)| \leq 1.
\end{eqnarray}
As in the prove of Lemma \ref{ll2},
\begin{eqnarray}
\sum_{k \in H_2} \frac{1}{P_k}|S_{2k}(c)| \leq 1.
\end{eqnarray}
By the above two facts,
\begin{eqnarray}
\frac{P_j - P_{ji}}{P_jP_{ij}} |S_{2i}(c)| + \frac{1}{P_j}|S_{2j}(c)| + \sum_{k \in H_2} \frac{1}{P_k}|S_{2k}(c)|\leq 1
\end{eqnarray}
Taking sum over all $c \in \mathcal C$, we get the desired result. (\ref{hmot}) and (\ref{hhai}) can be proved similarly.
\end{proof}

\begin{example}{\rm
Suppose that $(n,d,w) = (27,8,13)$. Consider $H_1 = \{11,12,13\}$. Let $i = 11, j = 12$. Then $m_{i,j} = 8 =d$. We have $H_2 = \{13\}$. If we let $k = 13$, then $m_{ik} = 6 < d$ and $m_{jk} = 4 < d$. We have
\begin{eqnarray}
\nonumber \begin{array}{lcrcl}
P_i & = & 26  &\geq&  T(11,13,11,14,8),\\
P_j & = & 1 &=& T(12,13,12,14,8),\\
P_{ij} & = & 20 &\geq& T(10,12,9,12,8),\\
P_{ji} & = & 1 &=&T(10,11,9,11,8),\\
P_k & = & 1 &=& T(13,13,13,14,8),
\end{array}
\end{eqnarray}
where the upper bounds
\begin{eqnarray}
\nonumber T(11,13,11,14,8) = T(2,13,3,14,8) \leq 26
\end{eqnarray}
and
\begin{eqnarray}
\nonumber T(10,12,9,12,8) = T(2,12,3,12,8) \leq 20
\end{eqnarray}
are taken from \cite{avz2}. Proposition \ref{dl7} shows that $A_{24} +
A_{26} \leq 1$ and $\frac{1}{26} A_{22} + \frac{6}{26} A_{24} +
A_{26} \leq 1$. The later inequality is equivalent to $A_{22} + 6
A_{24} + 26 A_{26} \leq 26$. Adding these two new constraints to
linear programming, we get
\begin{eqnarray}
\nonumber A(27,8,13) \leq 11897.
\end{eqnarray}
This improves
the upper of Agrell, Vardy, and Zeger: $A(27,8,13) \leq 11991$ (see
\cite{avz}), and the best known upper bound of Schrijver:
$A(27,8,13) \leq 11981$ (see \cite{s}).}
\end{example}

\section{More improvements on linear programming bounds}\label{ss4}

Another type of our constraints that can be added to the linear
programming is stated in the {\em $2$-row $k$-column formulas} in
this section. This type of constraints can in many cases help
decrease best known upper bounds on $A(n,d,w)$ by $1$.

Let $\mathcal C$ be an $(n,d,w)$ constant-weight code with distance distribution $\{A_{2i}\}_{i=d/2}^w$. Let $M$ be the number of codewords of $\mathcal C$. We consider $\mathcal C$ as a $M \times n$ matrix (where each $c \in \mathcal C$ is a row). For each $1 \leq i \leq n$, we let $x_i$ be the number of ones on the $i$th column of $\mathcal C$. For each $1 \leq k \leq n$, we define
\begin{eqnarray}
P^{-}_k(n;x) = \sum_{\substack{j=0 \\ j \mbox{ \footnotesize{odd} }}}^k \left(x \atop j \right) \left(n-x \atop k-j\right),
\end{eqnarray}
\begin{eqnarray}
P^{+}_k(n;x) = \sum_{\substack{j=0 \\ j \mbox{ \footnotesize{even} }}}^k \left(x \atop j \right) \left(n-x \atop k-j\right).
\end{eqnarray}
Hence, the Krawtchouk polynomial
\begin{eqnarray}
\nonumber P_k(n;x) & = & \sum_{j=0}^k (-1)^j \left(x \atop j \right) \left( n- x \atop k-j \right)\\
 &= &P^+_k(n;x) - P^-_k(n;x).
\end{eqnarray}

\begin{proposition} ($1$-row $k$-column formula). For each $k = 1,2, \ldots, M$,
\begin{eqnarray}
\sum_{\{u_1', \ldots, u_k'\}} wt(u_1' + \cdots + u_k') = M \cdot P^{-}_k(n;w),
\end{eqnarray}
where the sum is taken over all subsets $\{u_1', \ldots, u_k'\}$ containing distinct $k$ columns $u_1', \ldots, u_k'$ of $\mathcal C$.
\end{proposition}

\begin{proof}
Write $\mathcal C = (c_{ji})$. Let $S$ be the number of all $(c_{ji_1}, c_{ji_2}, \ldots, c_{ji_k})$ (these $k$ values are on the intersection of the row $j$ and the $k$ columns $i_1, i_2, \ldots, i_k$ of $\mathcal C$) such that $i_1 < i_2 < \cdots < i_k$ and such that $c_{ji_1} + c_{ji_2} + \cdots + c_{jk_k}$ is odd. Since each row of $\mathcal C$ has exactly $w$ ones, it will contribute
\begin{eqnarray}
P^{-}_k(n;w) = \sum_{\substack{j=0 \\ j \mbox{ \footnotesize{odd} }}}^k \left(w \atop j \right) \left(n-w \atop k-j\right)
\end{eqnarray}
to the number $S$. It follows that
\begin{eqnarray}
S = M \cdot P^{-}_k(n;w).
\end{eqnarray}
On the other hand, each $k$ columns $u_1', u_2', \ldots, u_k'$ of $\mathcal C$ contributes to $S$
\begin{eqnarray}
wt(u_1' + \cdots + u_k'),
\end{eqnarray}
which is the number of rows of $\mathcal C$ on which $u_1', u_2', \ldots, u_k'$ have an odd number of ones. Hence,
\begin{eqnarray}
S = \sum_{\{u_1', \ldots, u_k'\}} wt(u_1' + \cdots + u_k'),
\end{eqnarray}
where the sum is taken over all subsets $\{u_1', \ldots, u_k'\}$ containing distinct $k$ columns $u_1', \ldots, u_k'$ of $\mathcal C$. This finishes the proof of the proposition.
\end{proof}

The above $1$-row $k$-column formula plays an important role in the following $2$-row $k$-column formulas which will be added to linear programming to give improved upper bounds on $A(n,d,w)$.

\begin{proposition} ($2$-row $k$-column formulas). For each $k= 1, 2, \ldots, M$,
\begin{eqnarray}\label{congthuc2k1}
\nonumber \sum_{i=d/2}^w P_k^-(n;2i)A_{2i} \leq \frac{2}{M} \left[ \left( \left( n \atop k \right) - r_k\right)q_k(M-q_k) \right. \\
\left. + r_k (q_k +1)(M-q_k - 1)\right],
\end{eqnarray}
\begin{eqnarray}\label{congthuc2k2}
\nonumber -\sum_{i=d/2}^w P_k^+(n;2i)A_{2i} \leq  \frac{2}{M} \left[ \left( \left( n \atop k \right) - r_k\right)q_k(M-q_k) \right. \\
\left. + r_k (q_k +1)(M-q_k - 1)\right] - (M-1)\left(n \atop k\right),
\end{eqnarray}
where $q_k$ and $r_k$ are the quotient and the remainder, respectively, when dividing $M \cdot P_k^-(n;w)$ by $\left(n \atop k \right)$, i.e.
\begin{eqnarray}
M \cdot P_k^-(n;w) = q_k \left(n \atop k \right) + r_k,
\end{eqnarray}
with $0 \leq r_k < \left(n \atop k \right).$
\end{proposition}

\begin{proof}
Let $S$ be the number of all $2 \times k$ matrices
\begin{eqnarray}
A = \left(c_{ji_1} c_{ji_2} \cdots c_{ji_k} \atop c_{li_1} c_{li_2} \cdots c_{li_k} \right)
\end{eqnarray}
(the entries of $A$ are on the intersection of $2$ rows $j, l$ and $k$ columns $i_1, i_2, \ldots, i_k$ of the matrix $\mathcal C$) such that $j \not = l$, $i_1 < i_2 < \cdots < i_k$, and $A$ contains an odd number of ones. Two different rows $u$ and $v$ of $\mathcal C$ will contribute
\begin{eqnarray}
2 \sum_{\substack{j=0\\j \mbox{ \footnotesize{odd}}}}^k \left(d(u,v) \atop j\right) \left(n - d(u,v) \atop k-j \right) = 2 P^{-}_k(n;d(u,v))
\end{eqnarray}
to $S$. It follows that
\begin{eqnarray}\label{tt1}
S = \sum_{i = d/2}^w \sum_{\substack{u \not = v \in \mathcal C \\ d(u,v) = 2i }} P^{-}_k(n;2i) =M \cdot \sum_{i = d/2}^w P^{-}_k(n;2i)A_{2i}.
\end{eqnarray}
On the other hand, $k$ distinct columns $u_1', u_2', \ldots, u_k'$ of $\mathcal C$ will contribute
\begin{eqnarray}
2 \cdot wt(u_1' + \cdots + u_k') \cdot [M-wt(u_1' + \cdots + u_k')].
\end{eqnarray}
Hence,
\begin{eqnarray}\label{tt2}
S  = 2 \sum  wt(u_1' + \cdots + u_k')  \cdot [M-wt(u_1' + \cdots + u_k')],
\end{eqnarray}
where the sum is taken over all subsets $\{u_1', \ldots, u_k'\}$ containing distinct $k$ columns $u_1', \ldots, u_k'$ of $\mathcal C$. This sum contains $\left(n \atop k \right)$ summands. By the $1$-row $2$-column formula,
\begin{eqnarray}
\sum_{\{u_1', \ldots, u_k'\}} wt(u_1' + \cdots + u_k') = M \cdot P^{-}_k(n;w).
\end{eqnarray}
Since
\begin{eqnarray}
M \cdot P_k^-(n;w) = q_k \left(n \atop k \right) + r_k,
\end{eqnarray}
the right-hand side of (\ref{tt2}) is maximum, when $wt(u_1' + \cdots + u_k') = q_k$ in $\left(n \atop k \right) - r_k$ summands and $wt(u_1' + \cdots + u_k') = q_k + 1$ in the other $r_k$ summands.
It follows that
\begin{eqnarray} \label{tt3}
\nonumber S \leq 2 \left[ \left( \left( n \atop k \right)- r_k\right)q_k(M-q_k) \right. \\
 + \left. r_k (q_k +1)(M-q_k - 1)\right],
\end{eqnarray}
(\ref{tt1}) and (\ref{tt3}) give the desired formula (\ref{congthuc2k1}).

The proof of (\ref{congthuc2k2}) is similar. The only difference is that we count matrices $A$ having an even number of ones.
\end{proof}

\begin{example}{\rm
Suppose $(n,d,w) = (27,12,12)$. The best known upper bound for $A(27,12,12)$ is $A(27,12,12) \leq 140$. For $k = 1, 2,$ and $3$, the $2$-row $k$-column formula (\ref{congthuc2k1}) gives
\begin{eqnarray}
\nonumber 12A_{12}+14A_{14}+16A_{16}+18A_{18} + 20A_{20}+22A_{22}\\
\nonumber +24A_{24}  \leq  \frac{9333}{5},
\end{eqnarray}
\begin{eqnarray}
\nonumber 180A_{12}+182A_{14}+176A_{16}+162A_{18}+140A_{20}+110A_{22}\\
\nonumber +72A_{24} \leq \frac{859356}{35},
\end{eqnarray}
and
\begin{eqnarray}
\nonumber 1480A_{12}+1456A_{14}+1440A_{16}+1464A_{18}+1560A_{20}\\
\nonumber +1760A_{22}+2096A_{24} \leq 204715.
\end{eqnarray}
Adding these three constraints to the linear programming, we get
$$A(27,12,12) \leq \left\lfloor\frac{5604427}{40320} \right\rfloor + 1 = 139.$$
Therefore, $A(27,12,12) \leq 139$.
}
\end{example}

\begin{theorem}
For $n \leq 28$, improved upper bounds are summarized as follows (values in the parentheses are the best known upper bounds).
\begin{eqnarray}
\nonumber\begin{array}{lcrr}
\centerdot \quad A(18,6,8) & \leq & 427 & (428)\\
\centerdot \quad A(18,6,9) & \leq & 424 &(425)\\
\centerdot \quad A(20,6,10) & \leq & 1420 & (1421)\\
\centerdot \quad A(27,6,11) & \leq & 66078 &(66079)\\
\centerdot \quad A(27,6,12) & \leq & 84573 & (84574)\\
\centerdot \quad A(27,6,13) & \leq & 91079 &(91080)\\
\centerdot \quad A(28,6,11) & \leq & 104230 & (104231)\\
\centerdot \quad A(28,6,13) & \leq & 164219 & (164220)\\
\centerdot \quad A(28,6,14) & \leq & 169739 & (169740)\\
\hline
\centerdot \quad A(27,8,13) & \leq & 11897 &(11981)\\
\hline
\centerdot \quad A(24,10,10) & \leq & 170 &(171)\\
\centerdot \quad A(24,10,11) & \leq & 222 & (223)\\
\centerdot \quad A(24,10,12) & \leq & 246 & (247)\\
\centerdot \quad A(26,10,9) & \leq & 213 & (214)\\
\centerdot \quad A(27,10,9) & \leq & 298 &(299)\\
\centerdot \quad A(28,10,14) & \leq & 2628 & (2629)\\
\hline
\centerdot \quad A(26,12,10) & \leq & 47 &(48)\\
\centerdot \quad A(27,12,12) & \leq & 139 & (140)\\
\centerdot \quad A(27,12,13) & \leq & 155 & (156)\\
\centerdot \quad A(28,12,11) & \leq & 148&(149)\\
\centerdot \quad A(28,12,12) & \leq & 198 & (199)\\
\centerdot \quad A(28,12,13) & \leq & 244 &(245)\\
\centerdot \quad A(28,12,14) & \leq & 264 &(265).
\end{array}
\end{eqnarray}
\end{theorem}

\section{Applying to binary codes}\label{ss5}

An $(n,d)$ (binary) code is a subset $\mathcal C$ of
$\mathcal{F}^n$ such that distance between any two vectors in
$\mathcal C$ is at least $d$. The distance distribution
$\{A_i\}_{i=0}^n$ of $\mathcal C$ is defined by
\begin{equation}
A_{i} = \frac{1}{|\mathcal C|} \sum_{c \in \mathcal C} |S_{i}(c)|
\end{equation}
for $i = 0,1, \ldots, n$, where
\begin{eqnarray}
S_{i}(c) = \{u \in \mathcal C \mid d(u,c) = i\},
\end{eqnarray}
the set of all codewords $u$ in $\mathcal C$ at distance $i$ from
$c$.

Our technique in the previous section can be used to prove the
following theorem of Delsarte which makes linear programming
possible for codes.

\begin{theorem} \label{dtheorem} (Delsarte).
Let $\mathcal C$ be a code with distance distribution
$\{A_i\}_{i=0}^n$. Then
\begin{eqnarray}
\sum_{i=1}^n P_k(n;i) A_i \geq - \left(n \atop k \right)
\end{eqnarray}
for $k = 0,1, \ldots, n$.
If $M = |\mathcal C|$ is odd, then
\begin{eqnarray}
\sum_{i=1}^n P_k(n;i) A_i \geq \frac{1-M}{M}\left(n \atop k
\right).
\end{eqnarray}
\end{theorem}

\begin{proof}
Consider all $2 \times k$ matrices
\begin{eqnarray}
A = \left(c_{ji_1} c_{ji_2} \cdots c_{ji_k} \atop c_{li_1} c_{li_2} \cdots c_{li_k} \right)
\end{eqnarray}
(the entries of $A$ are on the intersection of $2$ rows $j, l$ and
$k$ columns $i_1, i_2, \ldots, i_k$ of the matrix $\mathcal C$)
such that $j \not = l$, $i_1 < i_2 < \cdots < i_k$. If $A$
contains an odd number of ones then we count $1$ for $S$ and if
$A$ contains an even number of ones then we count $-1$ for $S$.
Two different rows $u$ and $v$ of $\mathcal C$ will contribute
\begin{eqnarray}
\nonumber && 2 \displaystyle\sum_{\substack{j=0\\j \mbox{ \footnotesize{odd}}}}^k
\left(d(u,v) \atop j\right) \left(n - d(u,v) \atop k-j \right) \\
\nonumber &&  - \displaystyle2\sum_{\substack{j=0\\j \mbox{ \footnotesize{even}}}}^k
\left(d(u,v) \atop j\right) \left(n - d(u,v) \atop k-j \right)\\
\nonumber & = &   \displaystyle 2 P^{-}_k(n;d(u,v)) - 2P^+_k(n; d(u,v)) \\
& = &  - 2P_k(n;d(u,v))
\end{eqnarray}

to $S$. It follows that
\begin{eqnarray}\label{dthuc1}
S =  - \sum_{i = 1}^n \sum_{\substack{u \not = v \in \mathcal C \\
d(u,v) = i }} P_k(n;i) = - M \cdot \sum_{i = 1}^n P_k(n;i)A_{i}.
\end{eqnarray}
On the other hand, $k$ distinct columns $u_1', u_2', \ldots, u_k'$
of $\mathcal C$ will contribute
\begin{eqnarray}
\nonumber 2 \cdot wt(u_1' + \cdots + u_k') \cdot [M-wt(u_1' + \cdots + u_k')] \\
\nonumber - 2\left(wt(u'_1+\cdots+u_k') \atop 2 \right) \\
\nonumber  - 2\left(M-wt(u_1' + \cdots + u_k')\atop 2 \right)\\
\nonumber = 4 \cdot  wt(u'_1+\cdots+u'_k) \cdot [M-wt(u'_1+\cdots+u'_k)]\\
- M(M-1).
\end{eqnarray}
Hence,
\begin{eqnarray}\label{smlm}
\nonumber S  = 4 \sum wt(u_1' + \cdots + u_k')
 \cdot [M-wt(u_1' + \cdots + u_k')] \\
 - M(M-1)\left(n \atop k\right),
\end{eqnarray}
where the sum is taken over all subsets $\{u_1', \ldots, u_k'\}$
containing distinct $k$ columns $u_1', \ldots, u_k'$ of $\mathcal
C$. This sum is maximal when $wt(u'_1+\cdots+u'_k) = M -
wt(u'_1+\cdots+u'_k) = \displaystyle \frac{M}{2}$ for all $u'_1,
\ldots, u'_k$. It follows that
\begin{eqnarray}\label{dthuc2}
S \leq 4 \displaystyle \frac{M}{2} \frac{M}{2} \left(n \atop k
\right) -M(M-1) \left(n \atop k\right) = M \left(n \atop k\right).
\end{eqnarray}
(\ref{dthuc1}) and (\ref{dthuc2}) give
\begin{eqnarray}
\sum_{i=1}^n P_k(n;i) A_i \geq - \left(n \atop k \right).
\end{eqnarray}
However, if $M$ is odd then, the sum in (\ref{smlm}) is maximal when
\begin{eqnarray}
\nonumber & &wt(u'_1+\cdots+u'_k)[M-wt(u'_1+\cdots+u'_k)] \\
 &=& \displaystyle \frac{M-1}{2} \cdot \frac{M+1}{2}
\end{eqnarray}
for all $u'_1, \ldots, u'_k$.
Hence,
\begin{eqnarray}\label{dthuc3}
\nonumber S & \leq & 4 \displaystyle \frac{M-1}{2} \frac{M+1}{2} \left(n \atop k
\right) -M(M-1) \left(n \atop k\right) \\
& = & (M-1) \left(n \atop
k\right).
\end{eqnarray}
(\ref{dthuc1}) and (\ref{dthuc3}) give
\begin{eqnarray}
\sum_{i=1}^n P_k(n;i) A_i \geq \frac{1-M}{M} \left(n \atop k
\right).
\end{eqnarray}
\end{proof}

\begin{remark}
{\rm In the proof of Theorem \ref{dtheorem}, if we count the number of matrices $A$ having an odd number of ones and the number of matrices $A$ having an even number of ones separately, then we get the following inequalities. These inequalities are at least as good as the Delsarte inequalities because their sums give the Delsarte inequalities.
\begin{eqnarray}
\sum_{i=1}^n P_k^-(n;i)A_i \leq \frac{2M_1}{M} \left(n \atop k \right),
\end{eqnarray}
\begin{eqnarray}
-\sum_{i=1}^n P_k^+(n;i)A_i \leq -\frac{2M_2}{M} \left(n \atop k \right),
\end{eqnarray}
where $M_1$ and $M_2$ are given by
\begin{eqnarray}
M_1 = \left \{
\begin{array}{ll}
\displaystyle \frac{M^2}{4} & \mbox{ if } M \mbox{ is even }\\
\displaystyle \frac{M^2-1}{4} & \mbox{ if } M \mbox{ is odd }\\
\end{array}
\right.
\end{eqnarray}
and
\begin{eqnarray}
M_2 = \left \{
\begin{array}{ll}
\displaystyle \frac{M(M-2)}{4} & \mbox{ if } M \mbox{ is even }\\
\displaystyle \frac{(M-1)^2}{4} & \mbox{ if } M \mbox{ is odd }\\
\end{array}
\right..
\end{eqnarray}
}

\end{remark}

\ifCLASSOPTIONcaptionsoff
  \newpage
\fi

\end{document}